\documentclass[namedreferences]{solarphysics}

\pdfoutput=1

\usepackage[hyperref,optionalrh,showbiblabels]{spr-sola-addons} 
\usepackage{graphicx}        
\usepackage{color}           
\usepackage{breakurl}        

\chardef\us=`\_

\begin{document}

\begin{article}
\begin{opening}

\title{Decay of Activity Complexes, Formation of Unipolar Magnetic Regions and Coronal Holes in their Causal Relation}

\author[addressref={aff1},email={e-mail.golubeva@iszf.irk.ru}]{\inits{E.M.}\fnm{E.M.}~\lnm{Golubeva}}
\author[addressref=aff1,corref,email={e-mail.avm@iszf.irk.ru}]{\inits{A.V.}\fnm{A.V.}~\lnm{Mordvinov}}
\address[id=aff1]{Institute of Solar-Terrestrial Physics of  Siberian Branch of  Russian Academy of Sciences, Lermontov st., 126a, Irkutsk 664033, Russia }

\runningauthor{E.M.~Golubeva, A.V.~Mordvinov}
\runningtitle{Decay of Activity Complexes and Formation of Coronal Holes} 

\begin{abstract}
North--south asymmetry of sunspot activity resulted in an asynchro\-nous reversal of the Sun's polar fields in the current cycle. 
The asymmetry is also observed in the formation of polar coronal holes. 
A stable coronal hole was first formed at the South Pole, despite the later polar-field reversal there. 
The aim of this study is to understand processes making this situation possible. 
Synoptic magnetic maps from the Global Oscillation Network Group and corresponding coronal-hole maps from the Extreme ultraviolet Imaging Telescope aboard the Solar and Heliospheric Observatory and the Atmospheric Imaging Assembly aboard the Solar Dynamics Observatory are analyzed here to study a causal relationship between the decay of activity complexes, evolution of large-scale magnetic fields, and formation of coronal holes.
Ensembles of coronal holes associated with decaying active regions and activity complexes are presented. 
These ensembles take part in global rearrangements of the Sun's open magnetic flux. 
In particular, the south polar coronal hole was formed from an ensemble of coronal holes that originated after the decay of multiple activity complexes observed during 2014.

\end{abstract}


\end{opening}

\section{Introduction}
     \label{S-Intro} 

The Sun's magnetic activity exhibits a complicated spatio-temporal organization that is globally rearranged over an 11/22-year cycle. As a solar cycle progresses, the differential rotation transforms the poloidal field into a toroidal one, which emerges at the solar photosphere, forming active regions (ARs). Spatio-temporal distributions of emerging magnetic fluxes determine the special features of every 11-year cycle, including its North--South asymmetry. For example, the current cycle is characterized by highly asynchronous activity development in both hemispheres. To improve our understanding of the current cycle, it is of interest to study different manifestations of the Sun's magnetic activity in their causal relation.

ARs demonstrate a tendency to form activity complexes (ACs) in longitudinal ranges up to 60$^\circ$ with a lifetime of several solar rotations \citep{Gaizauskas1983}.
Analyses of the Greenwich Photoheliographic Results \citep{Brouwer1990} and the National Solar Observatory's Kitt Peak Vacuum Telescope magnetograms \citep{Harvey93} revealed large-scale patterns in magnetic-flux emergence.
\cite{Schijver2000} studied spatio-temporal behavior of activity nests in detail.
Using extended observational material, they concluded that about 50\,\% of bipolar regions are involved in activity nests.

After the decay of ACs, their magnetic fields are redistributed in the surrounding photosphere, forming unipolar magnetic regions (UMRs). Further evolution of UMRs is determined by the joint action of the differential rotation, meridional flows, and diffusion (\citealp{DeVore85}; \citealp{Wang89}). 
According to Babcock and Leighton's empirical concept, following polarity UMRs migrate poleward, forming the polar fields of the new cycle \citep{Leighton69}.

Taking into account the differential rotation, advection, and diffusion of photospheric magnetic fields, numerical simulations (\citealp{Baumann04}; \citealp{Schrijver08})  satisfactorily describe cyclic changes in solar activity. \cite{Jiang14} developed new models that consider magnetic-bipole tilts and reproduce in detail the cyclic reversals of polar magnetic fields. 
Based on the Babcock--Leighton mechanism, the flux-transport models improved our understanding of the cyclic evolution of the Sun's magnetic field (\citealp{Choudhuri95}; \citealp{Dikpati04}; \citealp{Kit11}; \citealp{Karak14}).

The current cycle is characterized by a noticeable North--South asymmetry in sunspot activity. Large ACs were observed in the northern hemisphere in 2011. The decay of these ACs resulted in the formation of following polarity UMRs. So, positive-polarity UMRs reached the North Pole and caused the change in the dominant polarity there by late 2012 -- early 2013 (\citealp{Mordvinov14}; \citealp{Sun15}; \citealp{Petrie15}). In the southern hemisphere, the highest level of activity was observed in 2014. The decay of long-lived ACs was accompanied by the formation of negative polarity UMRs, which reached the South Pole and caused the change of the dominant polarity there by the early 2015. Thus, the later activity development in the southern hemisphere led to the two-year delay in the magnetic field reversal at the South Pole.
 
The recent reversal of polar magnetic fields is accompanied by global rearrangement of the open magnetic flux. This rearrangement is also unusual. A stable coronal hole (CH) appeared first at the South Pole despite the later reversal there. To understand the early appearance of the polar coronal hole (PCH), it is important to study the evolution of CHs in their relation to large-scale solar magnetic fields and sunspot activity. Footpoints of low-latitude CHs usually appear near ARs  (\citealp{Cranmer09}; \citealp{Karachik10}; \citealp{Petrie13}). 
\cite{Bumba95} found cases of closer connection between low-latitude CHs, associated UMRs, and decaying ARs. They concluded that the phenomena involved are evolutionarily related.

Daily observations of CHs showed fast changes in the distribution of open flux \citep{Nolte78}. However, long-lived CHs are also observed. Large CHs appear within extended UMRs, which migrate from mid-latitudes to the Poles \citep{Harvey02}. Global rearrangements of open flux are accompanied by cyclic changes in the zonal structure of open and closed magnetic fields (\citealp{Wang09}; \citealp{Wang10}). 
These changes lead to a global rearrangement of open flux over the solar surface. The evolution of CHs is also determined by ARs, which interact with preexisting open flux that causes CH redistribution.  The interchange reconnection between open and closed magnetic-field lines makes open flux jump from one location to another (\citealp{Wang09}; \citealp{Wang10}). PCHs disappear before the polar field reversals, thereby designating the global rearrangement of solar open flux.

In this article, the causal relations between decaying ACs, their remnant magnetic fields and CHs are studied. The factors that affect the formation and evolution of the CHs are considered. 
Special attention is paid to the formation of the south PCH in the current cycle.

\section{Data and Method}
\label{S-aug}

The evolution of solar magnetic fields and changes in open flux are studied during 2006\,--\,2016. Synoptic maps produced by the {\it Global Oscillation Network Group} (GONG) for magnetic fields 
are analyzed for Carrington Rotations (CRs) 2047\,--\,2176. 
The synoptic magnetograms of moderate resolution  show clear large-scale magnetic fields.
The sizes of the maps are 360$\times$180 pixels (step in longitude is one degree, sine of latitude is uniformly sampled).
The GONG synoptic magnetograms  in the Ni~{\small{I}} 676.8 nm line show the photospheric radial component. The field strength varies from -1220 G to 1389 G during the period considered. 

  Despite their moderate spatial resolution, the maps represent magnetic-flux distributions in a universal format that is appropriate for studies of long-lived magnetic patterns in the solar atmosphere. 
  To identify large-scale patterns in magnetic flux distribution, it is efficient to average several successive synoptic maps. 
 We treat each of the maps as a possible state of magnetic flux that is represented in the course of its long-term evolution, whereas an averaged synoptic map represents an ensemble average, where long-lived patterns become visible.

The synoptic maps of CHs from the Solar and Heliospheric Observatory's Extreme ultraviolet Imaging Telescope (SOHO/EIT)  for CRs  1905\,--\,2123 and the Solar Dynamics Observatory's Atmospheric Imaging Assembly (SDO/AIA) for CRs 2124\,--\,2163
in the EUV 19.5 and 19.3 nm emission are available in FITS format at 
\url{lasco-www.nrl.navy.mil} and \url{secchi.nrl.navy.mil}, respectively.
We downsampled these maps to 360$\times$180 pixels. 
They show CH positions with values +1 or $-$1, according to a dominant magnetic polarity within a CH. Outside of CHs, the background values are equal to zero.

Synoptic maps of CHs for CRs 2164\,--\,2176 were derived from SDO/AIA daily images.
To trace CH boundaries we used the watershed segmentation technique (\citealp{Soille1999};   \citealp{Nieniewski02}). 
To avoid oversegmentation of solar EUV images, we used both preliminary- and post-processing.
After an appropriate smoothing and refinement, the resultant image detects CHs and low-intensity features in solar corona.
Similar approaches are also used for automatic tracking of CHs 
({\it e.g.}, \citealp{Barra09}; \citealp{Krista09}).

To study the polar-field reversal and global rearrangements in open magnetic flux, both data sets were analyzed in their causal relation. It is important that the averaging interval should be compatible with a lifetime of long-lived structures. Long-lived ACs recur at almost the same positions in the successive maps. In the averaged maps, ACs are identified as the domains of strong magnetic fields, which are typical for recurrent ARs. We average synoptic maps over five successive CRs, ascribing the resultant map to the medial CR (MCR). Such an averaging interval is about 0.4 years. This time is compatible with a typical AC lifetime.

The data should be corrected for differential rotation, when averaging. This corresponds to the Lagrangian description of magnetohydrodynamic systems taking into account the frozen-in property. In the study, the synoptic magnetograms are corrected according to the law of \cite{Howard90}:
\begin{equation}		\label{Eq-Om1}
{\mathrm{\omega = 14.33(\pm 0.054) - 2.12(\pm 0.35) sin^2 \varphi - 1.83(\pm 0.38) sin^4 \varphi}}.
\end{equation}
In the case of CH synoptic maps, the correction is made in accordance with the law of \cite{Timothy75}:
\begin{equation}     \label{Eq-Om2}
{\mathrm{\omega = 14.23(\pm 0.03) - 0.40(\pm 0.10) sin^2 \varphi}}.
\end{equation}
Here, $\varphi$  is the heliolatitude, and $\omega$ is the angular velocity expressed in deg day$^{\mathrm{-1}}$. 

The configuration of a neutral line at the source surface characterizes the general structure of the coronal magnetic field. These data were 
obtained from the PFSS model using classic boundary conditions (\citealp{Petrie13}; \citealp{Petrie14}).

\section{Solar-Activity Development and Evolution of PCHs at the End of Cycle 23}
\label{S-Evol}

Let us consider the solar-activity development at the end of Cycle 23, applying the method described.
 Long-lived ACs were observed in the southern hemisphere in 2006\,--\,2007, while sunspot activity in the northern hemisphere was very weak. 
 Small ARs emerged and decayed within the ACs.
 Then a prolonged minimum in solar activity was observed, without noticeable sunspots until mid-2009. 
   Such an activity development appeared to be favorable to study the decay of ACs  
 and their interaction with CHs in the southern hemisphere (\citealp{Wang10}; \citealp{deToma2011}; \citealp{Gesztelyi12}; \citealp{Petrie13}).

\begin{figure}    
   \centerline{\includegraphics[width=0.7\textwidth,clip=]{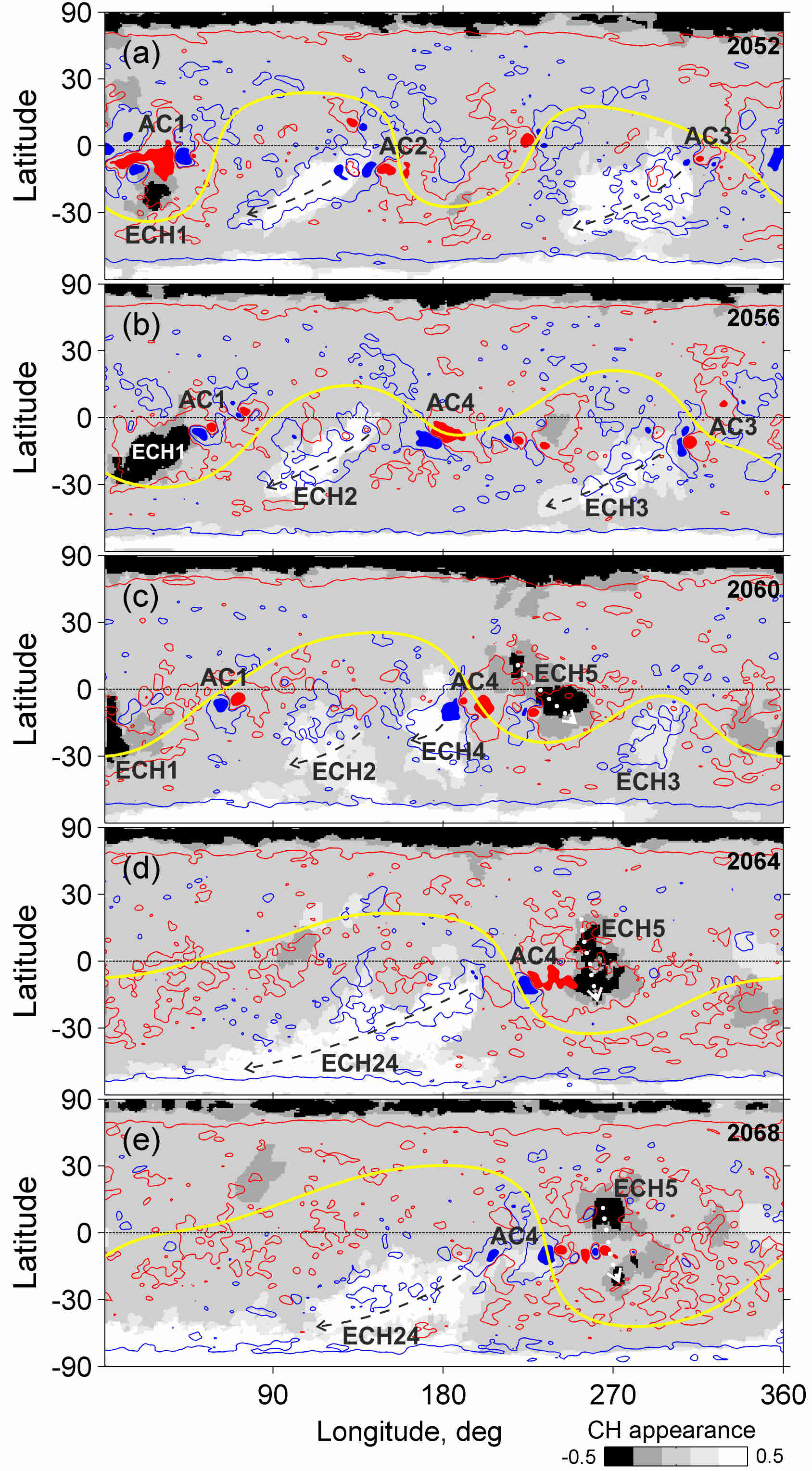}} 
   \caption{The averaged maps of CHs and magnetic fields  for MCRs 2052 (a), 2056 (b), 2060 (c), 2064 (d), and 2068 (e). 
 The CH appearance is shown according to the magnetic field polarities in black-to-white, when 
the absolute values exceeding 0.5 are equaled to $\pm$0.5 subject to their signs. 
 The blue contours correspond to magnetic field 1~G, and red contours to -1~G.
 The fields greater than 10~G are in blue, and the fields less than -10~G in red.
 The white dotted and black dashed arrows show the ECHs of leading and following polarity, respectively.
 The neutral lines are shown in yellow.}                 
   \label{F-cyc23}
   \end{figure}

Figure~\ref{F-cyc23}a shows the CH distribution (in grayscale) averaged over CRs 2050\,--\,2054. Here, the  MCR 2052 (8 January 2007 -- 4 February 2007) is  denoted.  All of the maps averaged are converted to the coordinate system of the map for the MCR, using Equation (\ref{Eq-Om2}). 
The averaged maps of CH appearance are shown in the limited range from -0.5 to $+$0.5 (black-to-white) for the image contrast. The absolute values greater than 0.5 are equaled to $\pm$0.5 according their signs. Thus, the positions of stable PCHs of negative/positive polarity, which are characterized by the values about $-$1/+1, are shown as black/white.

The magnetic-field distributions are averaged in the same manner, but applying the Equation (\ref{Eq-Om1}). The blue and red contours (Figure~\ref{F-cyc23}) correspond to magnetic field $\pm$1 G. 
ACs with more stronger fields (with absolute values greater than 10 G) are filled in blue and red. The neutral line on the source surface at 2.5 solar radii is yellow. Its position for all synoptic maps in an averaged set corresponds to the MCR. For example, Figure~\ref{F-cyc23}a  shows the neutral line corresponding to MCR 2052.

The coronal magnetic field was calculated in potential approximation with classical boundary conditions \citep{Petrie13}. It is characterized by a four-sector structure (Figure~\ref{F-cyc23}a). 
 In the southern hemisphere there are three open-flux domains where the low-latitude CHs appear.
 These domains are shown in black and white according to their polarities. 
Each of these open-flux domains is associated with AC1, AC2, and 
AC3. 
Averaging of the maps reveals the  frequent CH appearance domains
 which we call ensembles of coronal holes (ECHs). 

Further evolution of the large-scale magnetic structures can be seen in MCR 2056 
(27 April 2007 -- 25 May 2007, in Figure~\ref{F-cyc23}b), MCR 2060 (14 August 2007 -- 10 September 2007, in Figure~\ref{F-cyc23}c), MCR 2064 (1\,--\,29 December 2007, in Figure~\ref{F-cyc23}d), MCR 2068 (20 March 2008 -- 18 April 2008, in Figure~\ref{F-cyc23}e). 
 AC1 is associated with  low-latitude ECH1 which is embedded within the UMR of negative polarity (Figures~\ref{F-cyc23}a,b,c). 
 The following polarity ECHs are indicated with dashed arrows.
  After decay of the ACs, their ECHs  exist for several CRs. 
With the completing of AC3 decay, the coronal magnetic field becomes two-sectorial (Figures~\ref{F-cyc23}c\,--\,e).   
  
The sub-equatorial AC4 appears close to the neutral line and exists throughout a year (Figures~\ref{F-cyc23}b\,--\,e).
  AC4 seems to be west-shifted from map to map (Figures~\ref{F-cyc23}), in accordance with the differential rotation for sunspots.  
 AC4 appeared without an associated ECH (Figure~\ref{F-cyc23}b).
 Positive polarity ECH4 and negative polarity ECH5 appear by MCR 2060 
 (Figure~\ref{F-cyc23}c).
 Expanding ECH2 and ECH4 merge  to  ECH24  (Figures~\ref{F-cyc23}d,e).
 It is important to note the formation of the south PCH extension toward ECH24 
 (Figures~\ref{F-cyc23}d,e). 
  The interaction of small-scale and large-scale magnetic fields makes it evident 
 the cause--effect relationship between 
 ACs and the polar fields \citep{Wang04}.
 
With the development of  AC4, two open-flux domains are formed on opposite sides of the neutral line.
 Figure~\ref{F-cyc23}c shows the newly formed  ECH4 (positive) and ECH5 (negative).
It is reasonable to suggest that ECH4 and ECH5  are  linked  atop  the streamer belt.
They appear on different sides of the neutral line and are projected on peripheral parts of associated  AC4. 
\cite{Schrijver03} and \cite{Petrie13} noted that 
 footprints of the open flux can originate from  low-latitude CHs related to large ACs.

Our analysis of averaged synoptic maps reveals the ECHs associated with decaying ACs. A comparative analysis of the averaged synoptic maps of magnetic fields and CHs demonstrates the relationship between long-lived magnetic structures in the photosphere and corona.
 The neutral line goes through AC4 on the averaged map. The sharp bend in the neutral line is observed there. At the same time, the deviations of the heliospheric current sheet (HCS) from the Equator exceed 30$^\circ$.  

\begin{figure}    
   \centerline{\includegraphics[width=1\textwidth,clip=]{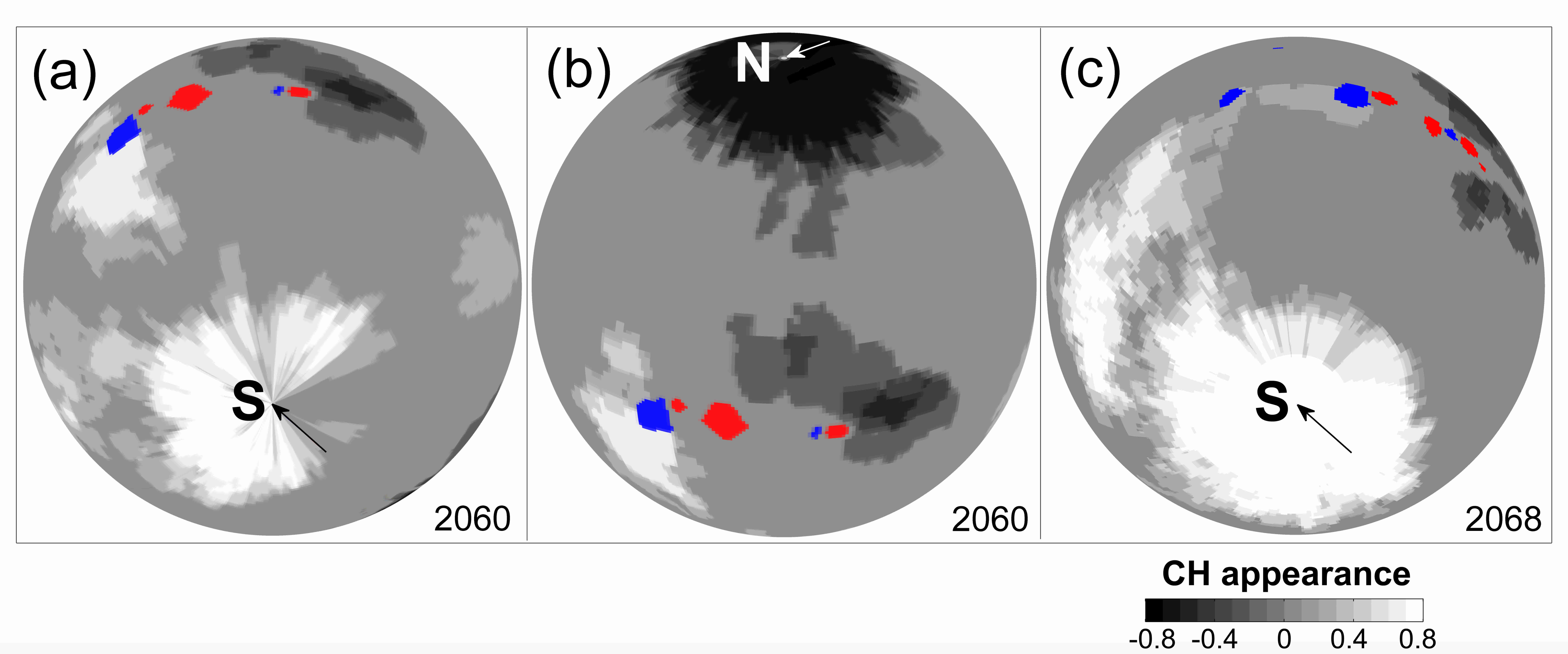} 
              }
\caption{The polar views derived from the 3D-reconstruction of the averaged maps: the South (S) and North (N) Poles for MCR 2060 (a and b) and the South Pole for MCR 2068 (c). 
 The CH appearance is shown according to the magnetic field polarities in black-to-white, when 
the absolute values exceeding 0.8 are equaled to $\pm$0.8 subject to their signs.
 The fields greater than 10~G are in blue, and the fields less than -10~G are in red.
 The Poles are marked with arrows directed along the zero meridian.}
   \label{F-pol23}
   \end{figure}

 The averaged synoptic map shown in Figure~\ref{F-cyc23}c was reconstructed to a 3D-object.
Figures~\ref{F-pol23}a,b show the polar projections with the decaying AC4 and associated open-flux domains for MCR 2060. 
 Both ECH4 and ECH5  are formed on opposite sides of AC4.
The southern and northern PCHs have  extensions towards AC4.
 The averaged distributions of magnetic fields and open-flux domains represent a new view of the Sun, when long-lived magnetic patterns become visible.

A similar polar view of the  synoptic map for MCR 2068 is shown in Figure~\ref{F-pol23}c.
Both the differential rotation and meridional flows force weak magnetic fields and ECH24 to form the helical structure that wraps around the south PCH. 
Further merger of ECH24 with PCH  results in its area increase. 
Thus, even at the end of the sunspot cycle, the decay of ACs led to the formation of ECHs that are transported to the Pole, thereby increasing the latitudinal extent of PCHs. 
Such a regularity was empirically established earlier ({\it e.g.}, \citealp{Tlatov14}), but the reasons were unknown.

\section{Reorganization of Open Magnetic Flux During the Polarity Reversal at the North Pole in the Current Cycle}
\label{S-ReorgN}

At the beginning of the current cycle, sunspot activity prevailed in the northern hemisphere, and the total sunspot area peaked there in late 2011. Decay of those ACs resulted in UMRs of predominantly following (positive) polarity. Their poleward transport by meridional flow caused the polar-field reversal at the North Pole in about 1.6 year (\citealp{Mordvinov14}; \citealp{Petrie15}). It was shown that the delayed development of the activity in the southern hemisphere led to a delay in the field reversal at the South Pole (\citealp{Sun15}; \citealp{Mordvinov16}). In the northern hemisphere, UMRs of both polarities migrated poleward with some predominance of the positive (following) polarity.

Analysis of bipolar sunspot group tilts showed that the leading-polarity UMRs were formed after the decay of groups where Joy's law was not obeyed 
 (\citealp{Yeates15}; \citealp{Mordvinov15}). 
The meridional transport of opposite magnetic polarities resulted in their mixture in the north polar zone  \citep{Mordvinov16}. The non-uniform magnetic field near the North Pole was unfavorable for the formation of a stable PCH until 2016.

\begin{figure}    
   \centerline{\includegraphics[width=0.7\textwidth,clip=]{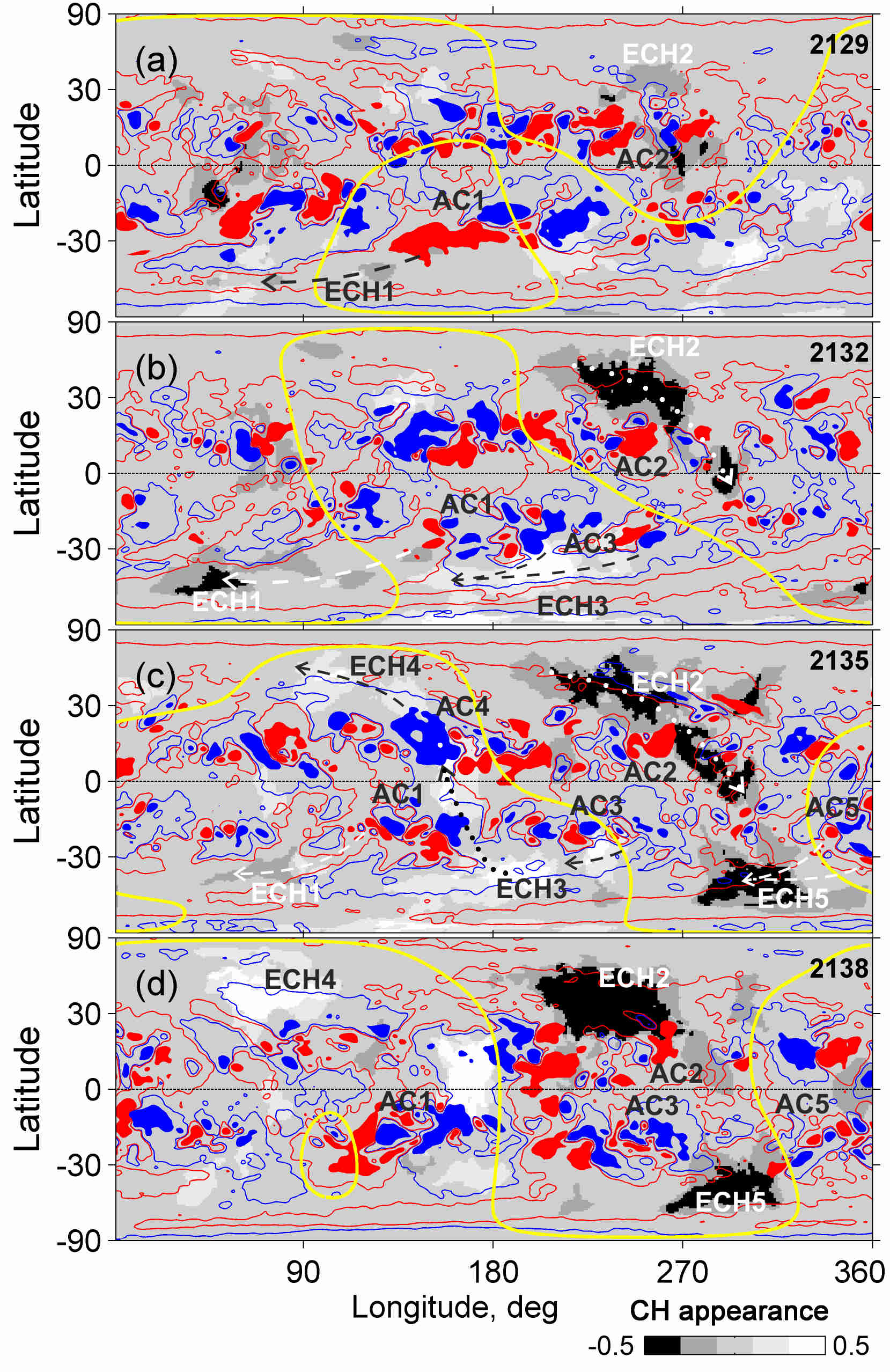} 
              }
  \caption{The averaged maps of CHs and magnetic fields  for MCRs 2129 (a), 2132 (b), 2135 (c), and 2138 (d).
 The CH appearance is shown according to the magnetic field polarities in black-to-white, when 
the absolute values exceeding 0.5 are equaled to $\pm$0.5 subject to their signs. 
 The blue contours correspond to magnetic field 1~G, and red contours to -1~G.
 The fields greater than 10~G are in blue, and the fields less than -10~G in red. 
 The white dotted arrows show the open-flux rearrangement.
 The white and black dashed arrows show the ECHs of following polarity.
 The neutral lines are shown in yellow.}
   \label{F-reN}
   \end{figure}

Figures~\ref{F-reN}a\,--\,d show the averaged distributions of CHs in accordance with their polarities (in grayscale) for MCR 2129 (8 October 2012 -- 4 November 2012),  
MCR 2132 (29 December 2012 -- 25 January 2013), 
MCR 2135 (21 March -- 17 April 2013), and
MCR 2138 (11 June 2013 -- 8 July 2013), respectively.  
The neutral lines are shown in yellow for MCRs 2129, 2132, 2135, and 2138. 
These MCRs are the temporal-averaged rotations in the averaging ranges considered. The numbers of the MCRs are indicated on each map at the top right. The neutral line shows large deviations from the Equator, and abrupt changes in magnetic flux are observed from MCR to MCR. Thus, the large-scale magnetic field is rearranged very rapidly. 
 Every averaged distribution of CHs or magnetic field is estimated over five original synoptic maps, including a medial CR and  previous and subsequent ones.
 The previous and subsequent maps are corrected for differential rotation (Equations (1),(2)).
 
The blue and red contours mark the weak fields (1 G) usually associated with long-lived UMRs, in accordance with the polarity signs. The areas of relatively strong fields (with absolute values greater than 10 G), which are associated with recurrent ARs and long-lived ACs, are marked by blue and red. The averaged magnetic-field distribution in MCR 2129 (Figure~\ref{F-reN}a) shows the prevalence of negative polarity at the North Pole, and positive polarity at the South Pole. 
 The northern PCH disappears although fragments of the southern PCH still remain.
Large deviations of the neutral lines  manifest the beginning of global rearrangement of the Sun's open flux.
Small ECHs are observed at mid- and low-latitudes.
There are multiple ARs and ACs in the northern and southern hemispheres. 
Despite their chaotic distribution several long-lived ACs play an important role in further evolution of the Sun's open flux. Figures~\ref{F-reN}a,b show such AC1 and AC2.

 Decaying AC1 forms the extended UMR of negative polarity.  
ECH1 is formed within the following polarity UMR that wraps around the South Pole, 
as is shown in Figures~\ref{F-reN}b\,--\,d where the dashed arrows mark the ECHs associated with the decaying ACs. 
  ECH2  of leading polarity originates near AC2 in the northern hemisphere (Figures~\ref{F-reN}a\,--\,d). 
After the northern PCH decay, its open flux relocates on the periphery of AC2.
The fast formation of ECH2 seems to occur because of the interaction between the polar open flux and AC2 magnetic field. 
Because of the interchange reconnection, the footpoints of open flux of negative polarity move from the North Pole to mid-latitude  AC2 (Figure~\ref{F-reN}b). 
Further ECH2 extension could be interpreted as multiple interchange reconnections that led to the jump-like ECH rearrangements. The transequatorial pattern of ECH2
indicates the transfer of open magnetic flux from the northern hemispheres to the southern one 
(Figure~\ref{F-reN}b,c). 
These open-flux rearrangements are shown with the white-dotted arrow.

One more open-flux domain of negative polarity formed at high latitudes of the southern hemisphere.
The decaying AC5 resulted in the ECH5 (Figure~\ref{F-reN}c,d). 
The ECHs of negative polarity ECH2 and ECH5 are arranged in such an order that specifies the direction of the global  rearrangement the Sun's open flux shown with a white-dotted arrow (Figure~\ref{F-reN}c).
The coronal magnetic field is characterized by a two-sector structure (Figure~\ref{F-reN}b,c,d). 
The open-flux domains ECH2 and ECH5 are within a sector of negative polarity (Figures~\ref{F-reN}).
 The open-flux domains  ECH3 and ECH4 appear within the sector of positive polarity (Figures~\ref{F-reN}c,d). At the bottom of this sector are fragments of the southern PCH.
The open-flux domains of positive polarity are also built in a chain that specifies the global rearrangement in opposite direction that is shown with the black dotted arrow.

 These changes can be interpreted as redistribution of open magnetic flux.
 The ECHs of opposite polarities  form  transequatorial structures, which indicates the transfer of open magnetic flux between the hemispheres (Figure~\ref{F-reN}c). Thus, the open-flux domains concentrate at mid- and low-latitudes after the disappearance of the PCHs. 
 It is likely that the open magnetic flux reconnects with closed magnetic bipoles during this time. Such fast changes in open flux could be caused by multiple interchange reconnections within both sectors (\citealp{Antiochos07}; \citealp{Edmondson09}; \citealp{Wang10}).
In this interpretation, some features look ambiguous. 
They depend on the  interplay between ACs and open-flux domains. 
Therefore, it is difficult to model the global rearrangements of the Sun's open flux  in detail \citep{Cranmer09}.

\section{Decay of ACs and Formation of the South PCH in the Current Cycle} 
\label{S-DecayS}

Analysis of time--latitude diagrams of averaged magnetic fields shows that the delayed development of the activity in the southern hemisphere results in a two-year delay in the polarity reversal at the South Pole during the current cycle (\citealp{Sun15}; \citealp{Mordvinov16}). 
Despite this delay, a stable PCH appears first at the South Pole. To understand one more peculiarity of the current cycle, let us consider the evolution of large-scale magnetic fields after the activity peak in 2014, when strong long-lived ACs were observed in the southern hemisphere of the Sun.

\begin{figure}    
   \centerline{\includegraphics[width=0.7\textwidth,clip=]{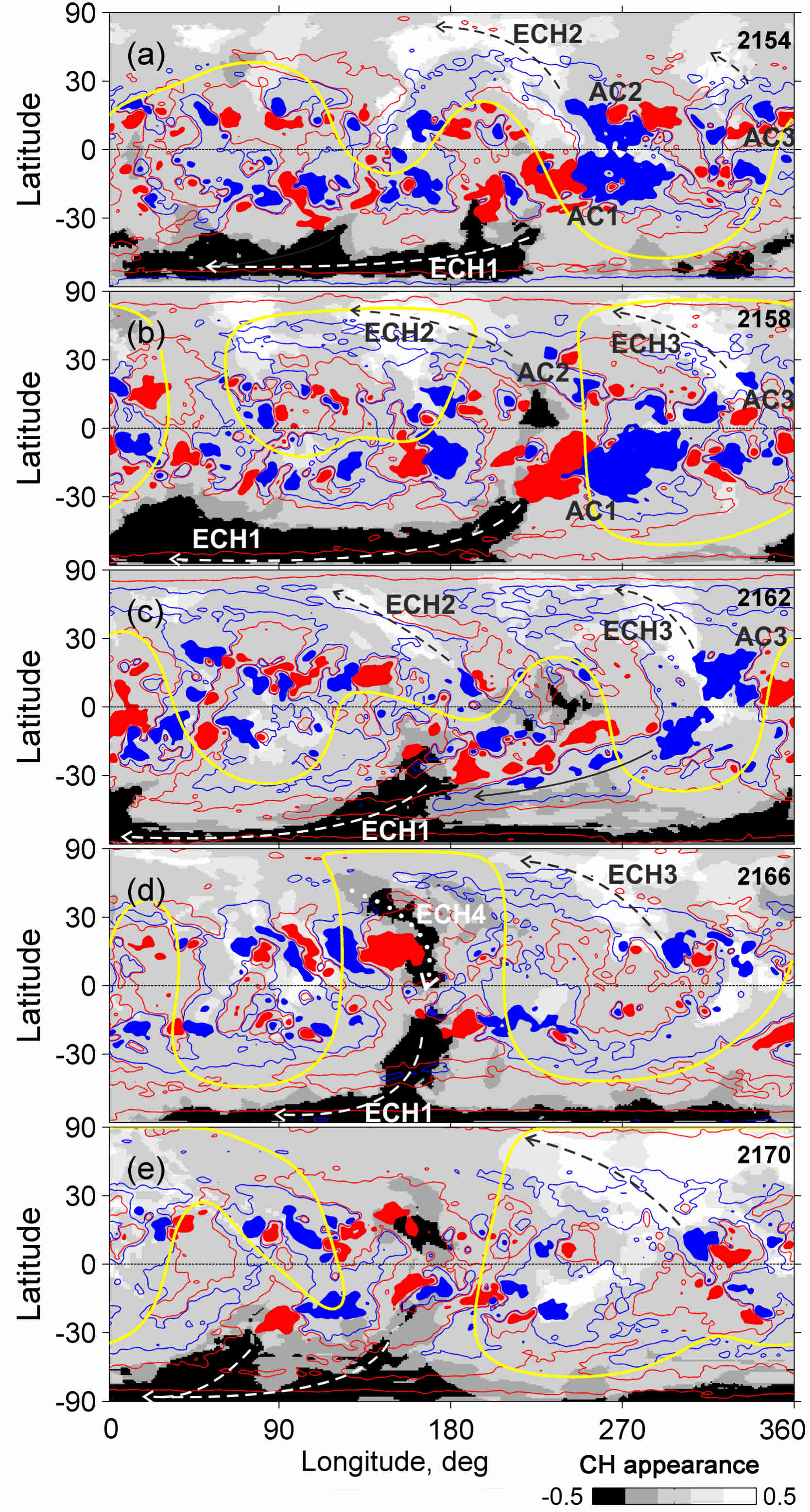} 
              }
  \caption{The averaged maps of CHs and magnetic fields for MCRs 2154 (a), 2158 (b), 2162 (c), 2166 (d), and 2170 (e).
 The CH appearance is shown according to the magnetic field polarities in black-to-white, when 
the absolute values exceeding 0.5 are equaled to $\pm$0.5 subject to their signs. 
 The blue contours correspond to magnetic field 1~G, and red contours to -1~G.
 The fields greater than 10~G are in blue, and the fields less than -10~G in red.
 The white and black dashed  arrows indicate the following polarity ECHs.
 The white dotted arrow shows the open-flux rearrangement of leading polarity.
 The black solid arrow indicates the leading polarity UMR.
 The neutral lines are shown in yellow.
 }                                                                   
   \label{F-reS}
   \end{figure}

Figures~\ref{F-reS} show the sequence of maps averaged over five adjacent CRs. The distributions of CH appearance are presented with a correction for differential rotation (Equation (2)).
Weak magnetic fields ($\pm$ 1 G) are marked as blue and red contours in accordance with their polarity signs.
The blue and red areas correspond to relatively strong fields (with absolute values greater than 10 G), which are associated with  long-lived ACs. The yellow lines correspond to the neutral line in MCRs 2154, 2158, 2162, 2166, and 2170.

Figures~\ref{F-reS}a,b present the averaged distributions for MCR 2154 (21 August 2014 -- 17 September 2014) and for MCR 2158 (8 December 2014 -- 4 January 2015). The huge areas of stronger fields indicate multiple ACs that prevail in the southern hemisphere. The giant AC1 covers the longitude range 220$^\circ$--300$^\circ$. The most powerful ARs in the cycle were observed in AC1. 
As the AC1 decays, 
an extensive UMR of negative polarity is formed. A lot of CHs are observed within the bounds of the UMR, forming the ensemble (ECH1).
	
ECH1 details near the AC1 are characterized by their steeper tilts relative to the Equator (Figure~\ref{F-reS}a,b). These features show their recent appearance due to the newly formed CHs. 
The inclined structures, which are associated with the negative polarity UMRs, are observed in MCR 2158 (Figure~\ref{F-reS}b) and MCR 2162 (Figure~\ref{F-reS}c, 28 March 2015 -- 24 April 2015). Further evolution of UMRs is caused by diffusion, differential rotation, and meridional transport.
 Structural changes in ECH1 take place. The subsequent maps (Figure~\ref{F-reS}c,d,e) show the merger of high-latitude UMRs that results in the formation of the ringed structure around the South Pole. Eventually, the UMRs reach the South Pole and cause the polarity reversal there. 
 In parallel with the UMR evolution, the ECH1 demonstrates similar rearrangements. These changes results in the PCH formation at the South Pole.

 Several  ARs, as well as long-lived AC2 and AC3 are observed in the northern hemisphere (Figures~\ref{F-reS}).  
Decaying AC2 and AC3 produce positive polarity UMRs  
 and associated ECHs. 
In Figures~\ref{F-reS} these ECHs are marked by the arrows  ECH2 and ECH3.
The UMRs of opposite polarities    are also observed     at the high latitudes. 
There is no stable PCH at the North Pole by mid-2016.
The open-flux domain appears suddenly within the negative polarity UMR (Figure~\ref{F-reS}d, 14 July 2015 -- 11 August 2015). It is an example of a leading polarity ECH that is marked by the dotted arrow ECH4. 
Its rapid appearance and the further rearrangements 
(Figure~\ref{F-reS}e, 1\.--\.28 November 2015)
demonstrate  the open-flux transfer between the hemisphere and enlargement of the south PCH.

The AC2 connects with AC1 via a common transequatorial area of positive polarity. 
In addition, the following polarity UMR in the northern hemisphere, there is its ``anomalous'' branch which reaches mid-latitudes in the southern hemisphere (the solid arrow in Figure~\ref{F-reS}c). 
The nature of this anomaly is usually associated with violations of Joy's law (\citealp{Yeates15}; \citealp{Mordvinov15}; \citealp{Mordvinov16}). The further intrusion of the leading-polarity UMR resulted in a narrow gap in the south ECH where CH occurrence reduced (the solid arrow in Figure~\ref{F-reS}c). 
 The cancellation of the opposite polarities results in the PCH area decrease for MCR 2166 and 2170 (Figure~\ref{F-reS}d and Figure~\ref{F-reS}e).
 The cancellation  in the Sun's polar regions is also accompanied by the interaction of open and closed magnetic configurations \citep{Edmondson09}.
 
\begin{figure}    
   \centerline{\includegraphics[width=1\textwidth,clip=]{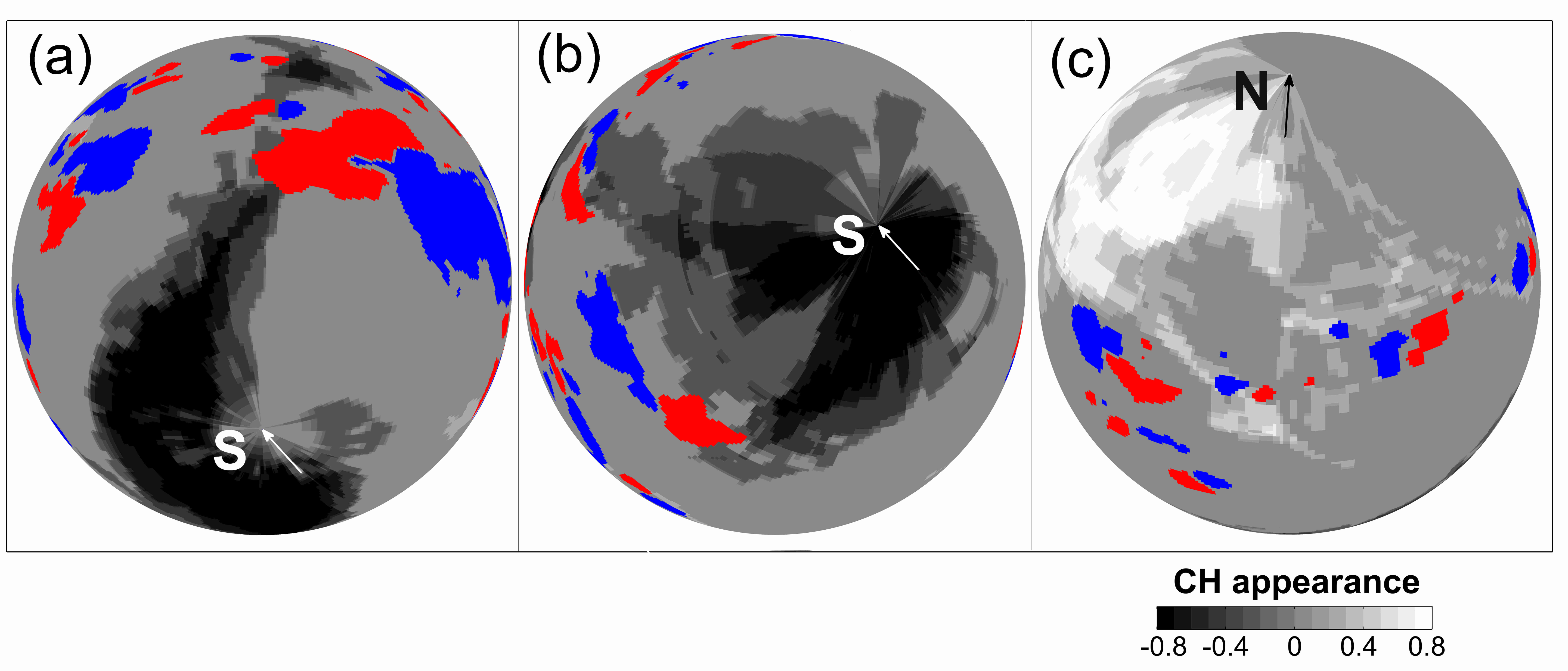} 
              }
    \caption{The south polar view of the averaged map for MCR 2158 (a) and the south and north polar views of the averaged map for MCR 2170 (b and c).  
 The CH appearance is shown according to the magnetic field polarities in black-to-white, when 
the absolute values exceeding 0.8 are equaled to $\pm$0.8 subject to their signs.
 The fields greater than 10~G are in blue, and the fields less than -10~G are in red.
 The North (N) and South (S) Poles are marked with arrows directed along the zero meridian.}   
   \label{F-poS}
   \end{figure}

 The 3D-reconstruction of the averaged map for MCR 2158 is shown in Figure~\ref{F-poS}a. 
What we see here is the south PCH formation due to  decaying AC1. 
The 3D-reconstruction of the map for MCR 2170 shows the further development of the south PCH and the early formation of the north PCH. 
Figures~\ref{F-poS}b,c show the Sun's views from the southern and northern poles, respectively.
The south PCH 
  looks highly asymmetric.
 At the same time its formation continues due to the ECHs associated with
 the adjacent decaying ACs (Figure~\ref{F-reS}e). 
 
\section{Conclusions}
\label{S-Conc}

 The analysis of averaged maps of the magnetic fields and solar EUV emission reveals the domains of frequent CH appearance related to decaying  ACs.  
The resulting ECHs and embedded open-flux domains exist for several months.
Some ECHs slowly evolve within the UMRs of the following polarities (in terms of Hale's law). 
The PCHs are formed due to the high-latitude ECH redistributions and their poleward transport within the bounds of associated UMRs. 
We demonstrated that areas of PCHs increase towards the end of the solar cycle. 
So, they are largest at solar minima.

A comparative study of the maps also reveals the interactions between PCHs and ECHs associated with the decaying ACs. The multiple extensions from the PCHs reach the ECHs of the same polarities. Thereby they enlarge the open-magnetic-flux areas in the corona that determine the general configurations of the global neutral line and HCS.

 The leading polarity ECHs originate because of fast  relocation of high-latitude open flux towards the adjacent ARs and ACs. Such ECHs sometimes show jump-like changes because of the interchange reconnection of open and closed magnetic fields. Some newly formed ECHs of opposite magnetic polarities are possibly linked through upper magnetic loops of embedded streamer belt.

During the polar-field reversals, the open flux from the PCHs is redistributed through the ECHs associated with the decaying ACs. Further evolution of the ECHs results in their transequatorial patterns which indicate the open-flux transfer between the northern and southern hemispheres. The mechanism of interchange reconnection between the open and closed magnetic fields is appropriate to describe the fast rearrangements in the open flux during this period.

 Our analysis reveals a causation between the decaying ACs and the associated UMRs, as well as the domains of frequent CH appearance. 
This relationship is of great importance for understanding of CH evolution. 
In particular, the decay of long-lived ACs observed in 2014 resulted in the extended ECH that was transported poleward to form the south PCH.

%

\begin{acks}
  This work utilizes data
obtained by the Global Oscillation Network Group (GONG) program, managed by the National
Solar Observatory, which is operated by AURA, Inc. under a cooperative agreement with the
National Science Foundation. The data were acquired by instruments operated by the Big Bear Solar Observatory, High Altitude Observatory, Learmonth Solar Observatory, Udaipur Solar
Observatory, Instituto de Astrof\'ysica de Canarias, and Cerro Tololo Interamerican Observatory.
Synoptic magnetograms from Baikal Solar Observatory were also used for the comparative analysis of the polar-field reversals.
The authors also employed synoptic maps of the solar EUV emission prepared by SOHO/EIT science team and NASA/SDO and the AIA, EVE, and HMI science teams.

This work was supported by the project II.16.3.1 under the Program of Fundamental Research of SB RAS.
The authors are grateful to the referee for the useful suggestions.

\end{acks}

\section{Disclosure of Potential Conflicts of Interest}
The authors declare that they have no conflicts of interest.



\bibliographystyle{spr-mp-sola}
\bibliography{GolubevaMordvinov3}  

\IfFileExists{\jobname.bbl}{} {\typeout{}
\typeout{****************************************************}
\typeout{****************************************************}
\typeout{** Please run "bibtex \jobname" to obtain} \typeout{**
the bibliography and then re-run LaTeX} \typeout{** twice to fix
the references !}
\typeout{****************************************************}
\typeout{****************************************************}
\typeout{}}

\end{article} 

\end{document}